# Reconstructing human activities via coupling mobile phone data with location-based social networks

Le Huang [a,b,1], Fan Xia [c,1], Hui Chen [c], Bowen Hu [d], Xiao Zhou [f], Chunxiao Li [e,*], Yaohui Jin [a,b,*], Yanyan Xu [a,b,*]

[a] MoE Key Laboratory of Artificial Intelligence, AI Institute, Shanghai Jiao Tong University, Shanghai 200240, China
[b] Data-Driven Management Decision Making Lab, Shanghai Jiao Tong University, Shanghai 200030, China
[c] School of Information Science and Technology, Shandong University, Qingdao 266237, China
[d] Technology R&D Department, Data Intelligence Division, China Unicom Digital Technology Co., Ltd., Beijing 100166, China
[e] Antai College of Economics and Management, Shanghai Jiao Tong University, Shanghai 200030, China
[f] Gaoling School of Artificial Intelligence, Renmin University of China, Beijing 100872, China

## ABSTRACT

In the era of big data, the ubiquity of location-aware portable devices provides an unprecedented opportunity to understand inhabitants' behavior and their interactions with the built environments. Among the widely used data resources, mobile phone data is the one passively collected and has the largest coverage in the population. However, mobile operators cannot pinpoint one user within meters, leading to the difficulties in activity inference. To that end, we propose a data analysis framework to identify user's activity via coupling the mobile phone data with location-based social networks (LBSN) data. The two datasets are integrated into a Bayesian inference module, considering people's circadian rhythms in both time and space. Specifically, the framework considers the pattern of arrival time to each type of facility and the spatial distribution of facilities. The former can be observed from the LBSN Data and the latter is provided by the points of interest (POIs) dataset. Taking Shanghai as an example, we reconstruct the activity chains of 1,000,000 active mobile phone users and analyze the temporal and spatial characteristics of each activity type. We assess the results with some official surveys and a real-world check-in dataset collected in Shanghai, indicating that the proposed method can capture and analyze human activities effectively. Next, we cluster users' inferred activity chains with a topic model to understand the behavior of different groups of users. This data analysis framework provides an example of reconstructing and understanding the activity of the population at an urban scale with big data fusion.

## 1. Introduction

As the population explodes in urban areas, people are increasingly aware of the complexity and importance of human mobility. Specifically, human mobility data is of great significance in assisting urban planning (Batty et al., 2012; Hu et al., 2019; Xu et al., 2020; Jiang et al., 2016), traffic control and travel behavior management (Tu et al., 2020; Pan et al., 2013; Xu and González, 2017; Xu et al., 2018), social governance (De Nadai et al., 2020; Xu et al., 2019), and discovering the influence of natural disasters (Yin et al., 2020), etc. Modeling of human mobility also plays a crucial role in the control of COVID-19, from the initial spreading stage to the economic reopening stages (Chang et al.,

2020; Kraemer et al., 2020; Mistry et al., 2021; Aleta et al., 2020). The traditional collection methods are completed through questionnaire surveys. These methods are only applicable to small samples. In addition to being difficult to update, they are time-consuming and costly. Moreover, the questionnaire surveys are usually collected by users recalling their trips on the day, so the specific time of the trip is not accurate and some unobtrusive small trips may likely be missed (McDonald, 2008). With the development of information and communication technologies (ICTs), the ubiquity of mobile devices and location-based services have provided more channels for the collection of individual trajectories, such as geo-information on social networks (Liu et al., 2021), location data in mobile phones (Sagl et al., 2014, 2016, 2013, 2016), GPS trajectories (Siła-Nowicka et al., 2016), smart public transportation cards (Long et al., 2012), credit cards (Clemente et al., 2018), and taxis (Xu et al., 2021; Hu et al., 2021). Long et al. combined the bus smart card data and the land use map to analyze the jobs-housing places and commuting patterns (Long et al., 2012). Hu et al. modeled the relationship between urban functional structure and traffic spatial interaction to classify urban functions at the road segment level (Hu et al., 2021).

* Corresponding authors.
  *E-mail addresses:* chunxiao@sjtu.edu.cn (C. Li), jinyh@sjtu.edu.cn (Y. Jin), yanyanxu@sjtu.edu.cn (Y. Xu).
  [1] These authors contributed equally in this work.

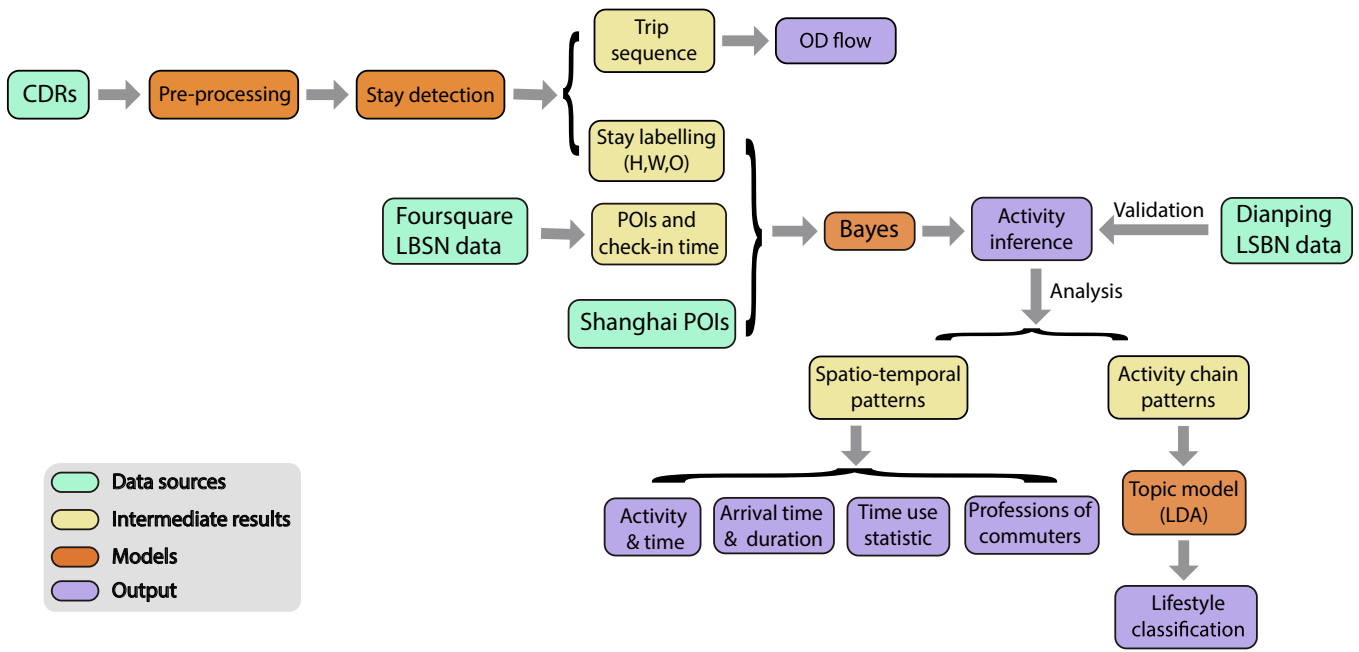

**Fig. 1. Pipeline of the proposed data analysis framework.** Cells in green refer to the original input data; cells in yellow refer to the intermediate results; cells in orange refer to the methods, including the stay detection, the Bayesian inference and so on.

Each type of data has its own shortcomings. The social network data are actively shared by the users and they do not always check in at every visited location. That is, the social network data can not collect the continuous sequence of individual mobility behavior. In addition, social network data can not uniformly cover the population due to the various popularity of location-aware applications in different ages. In contrast, mobile phone data can record a user's continuous movement behavior but the recorded locations are derived from the interacted base stations rather than the actual location. The covering radii of base stations range from 50 meters in crowded urban regions to about 300 meters in rural areas, indicating that we can not directly identify the users' activity from the mobile phone data.

Targeting inferring the activity type of each individual mobile phone user, we desire to consider not only the possible activities in the area covered by the interacted base station but also the pattern of the attractiveness of different types of facilities. To this end, we couple mobile phone data with two other datasets, POI data, and LBSN data. The POI dataset provides the longitude, latitude, and type of each facility, implying the association between the semantic and spatial information. The LBSN dataset provides the attractive flow of each type of facility every timeslot (e.g., 15 min, 1 h), implying the association between the semantic and temporal information. Through integrating these datasets, we are able to consider both temporal and spatial constraints. To this end, we propose a Bayesian inference model to assign a possibility to each activity type given a user's arrival time and destination. Taking Shanghai as an example, we leverage the anonymized mobile phone data of 1,000,000 residents during two weeks. The mobile operator records not only the call detail but also the usage of internet data, namely XDRs. For each user, we first detect significant stay locations via clustering the records in space with her raw XDR traces, and next label these stay locations as *home*, *work* (if available), or *other*. Then we devise a Bayesian model to reconstruct the activity chains via coupling the stay locations with the POI information and visitation pattern, extracting from the LBSN data. The reconstruction results are compared with a real-world dataset collected by Dianping (Dianping, 2022), an online platform for discovering local businesses and services in China. The validation results show that our model can effectively capture the laws of human activities on a large scale. After that, we analyze the spatio-temporal patterns of their activity behavior. Compared with the American Time Use Survey (ATUS), we find that the inhabitants in Shanghai devote more time to work than Americans and they have more involvement in activities at home than outdoors. We also use a topic model to classify the users into six groups based on the patterns of their inferred activity chains and analyze their behavior characteristics of each type. Overall, the key contribution of our study is the new data analysis framework that can assist in modeling human activity patterns at an urban scale with big data fusion, providing the basis for developing smart cities. The complete data analysis framework is presented in Fig. 1, including data sources in green cells, intermediate results in yellow cells, key algorithms in orange cells, and results in purple cells.

## 2. Related work

### 2.1. Social network data and mobile phone data analysis

In the information era, with the popularity of location-based services on personal mobile phones, people can share their current Location anytime and anywhere, forming location-based social networks (LBSN). Typical LBSN data, including Foursquare and Dianping, record the anonymized user ID, timestamp, location, and the visited facility in each item, providing an opportunity to understand the interaction between the population and the built environments (Huang and Wong, 2016; Tu et al., 2017; Hasan et al., 2013; Calafiore et al., 2021; Gao et al., 2017). Using the Twitter data for Washington, DC, Huang et al. introduced an approach to analyzing the activity patterns with different socioeconomic statuses (Huang and Wong, 2016). Via combining the Foursquare and Twitter data, Hasan et al. found that people tend to select places with diminishing probability following a Zipf's law (Hasan et al., 2013). Using Foursquare to provide insight into the urban structure, Calafiore et al. create a framework for the identification of urban functions (Calafiore et al., 2021). As mentioned before, the social networks data can only represent the behavior of a group of the population who frequently use social media platforms.

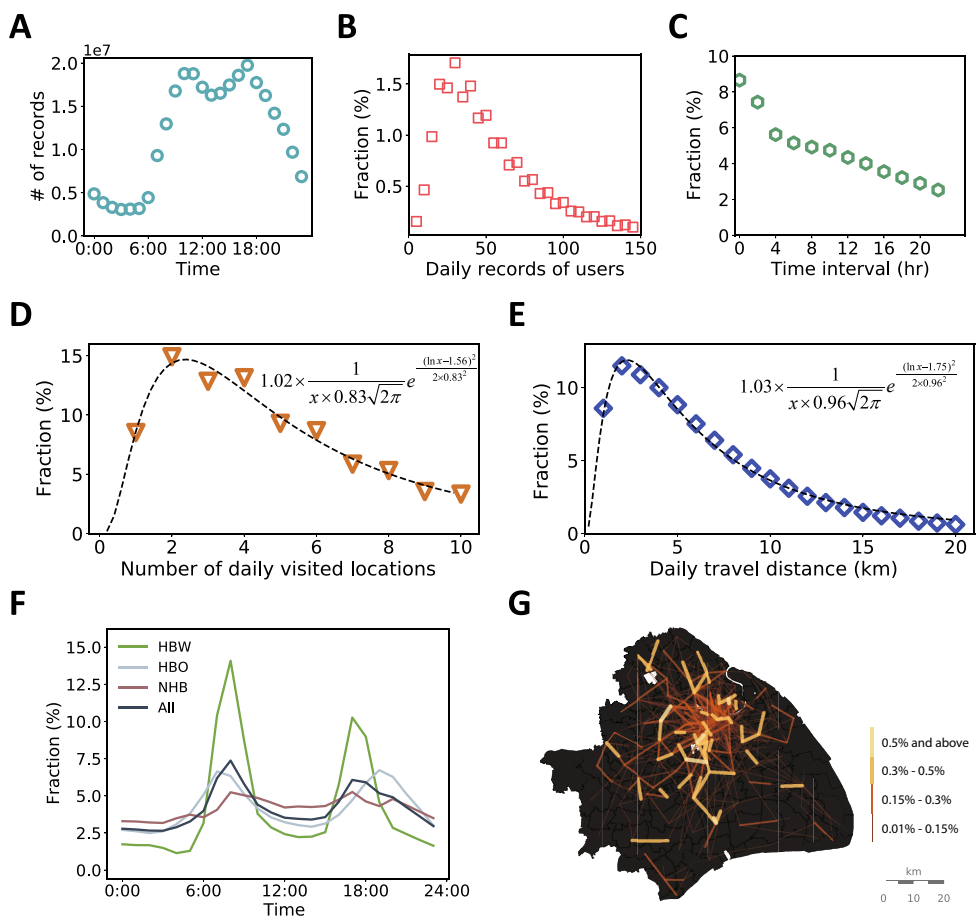

**Fig. 2. Statistic characteristics of the XDRs data and the estimated travel demand.** (A) Number of records of raw data in each hour during one weekday. (B) Distribution of the number of records per user per day. (C) Distribution of the time interval between two consecutive records of the same user. (D) Distribution of the number of daily visited locations, following a log-normal distribution. Mobile phone users visit nearly 5 locations per day on average. (E) Distribution of the number of daily travel distances, following a log-normal distribution. The average daily travel distance is 5.75 km in Shanghai. (F) Fraction of travel demands of different trip purposes by the hour including home-based work (HBW), home-based other (HBO), and non-home based (NHB) on weekdays. HBW trips display clearly peaks during the morning and evening commuting hours. (G) Visualization of travel flow during the morning peak hours on one typical weekday.

The mobile phone data, e.g., call detail records (CDRs) and x-detail records (XDRs), record the base stations used by the phone users when they are making phone calls, sending or receiving messages, or using the data-channel, and thus pinpoint users with the locations of base stations. The mobile phone data are passively collected by the mobile operators and cover a very large proportion of the population. There are a number of examples analyzing mobile phone data to study human mobility (Jiang et al., 2016; Sevtsuk and Ratti, 2010), daily rhythms (Monsivais et al., 2017; Roy et al., YYYY), social network analysis (Fudolig et al., 2020; Fudolig et al., 2021) and urban spatial structure (Louail et al., 2015; Lenormand et al., 2015). Jiang et al. presented a mechanistic modeling framework to analyze the individual daily mobility including stay duration and daily mobility motif distribution (Jiang et al., 2016). Via studying the mobile calling activity, Monsivais et al. found that the length and timings of urban daily rhythms sensitively depend on the seasonal changes of sunlight (Monsivais et al., 2017). Sagl et al. employed the self-organizing map (SOM) approach to promote the exploration of collective human activity, such as the spatial variations in intensity and similarity (Sagl et al., 2014).

## 2.2. Human activities

Mobile phone data is one of the largest data resources depicting human mobility behavior in terms of its coverage, but it presents challenges when attempting to infer user activity because of the coarse localization of individual users. To curb this challenge, a naive solution is to randomly select one facility or point of interest (POI) from those located in the Voronoi polygon of the interacted base station. Such a method neglects the heterogeneous visitation patterns of different types of facilities. For instance, restaurants always attract large customer flows during the noon and evening peak hours, while the shopping malls attract more customers on weekends and holidays. Therefore, there is a must to consider the attractiveness of different types of facilities. Xie et al. proposed to infer the activity type by associating the activity area with POIs according to the distance between them (Xie et al., 2009). Researchers assumed that the nearer the POI locates to the user, the larger probability of the facility the user engaged in (Wang (2012), Zhao et al. (2015). Besides, researchers also used Lévy model (Rhee et al., 2011) or gravity model (Goh et al., 2012) to infer the probability of users accessing various types of POIs. On the other hand, time constraints also play important roles in inhabitants' activities and have been combined with space constraints to infer individuals' activity types (Aslam et al., 2020; Zhao et al., 2020). For example, if the arrival time is beyond the opening hours of a shopping mall, the visit probability of the mall will be nearly zero. But this requires the complete opening hours information about the POI. Moreover, the rhythm of daily life also impacts the attractiveness of each type of POI (Spinsanti et al., 2010). Gong et al. inferred the activity types based on the variations in probability of population's activity within a day (Gong et al., 2016). However, for the mobile phone data, we can only localize the users in the area surrounding the base station, which covers various types of facilities. These distance-based methods can not be directly used to infer individuals' activity types. The surrounding environment is also an important factor in detecting activity types. Recent methods infer the activity of users with their trajectory similarities to the ones with known activities (Liu et al., 2021; Cai et al., 2016). Such methods require a large number of labeled trajectories to achieve satisfying performance. In addition, researchers have introduced data analyzing models to understand the activity types. For instance, the principal component analysis (PCA) is used to identify the main components of daily activities (Eagle and Pentland, 2009). In recent years, the latent Dirichlet allocation (LDA) model has also played a great role in inferring activity patterns (Zhao et al., 2020; Zhao et al., 2018) and identifying major crowd behaviors that recur over time (Ferrari et al., 2011).

**Table 1**
Mapping of the activity types, primary POIs, Foursquare and ATUS.

| Activity types | Primary POIs | Foursquare | ATUS |
|---|---|---|---|
| Shopping | Shopping, Car service | Store, Bookstore, …, Bookstore, Shop | Purchasing goods and service |
| Daily life | Banks, Hospitals, Religion | Office, Government Building, bank, Church, …, Medical Center, Temple | Organizational, Civic, Religious activities |
| Transport | Transportation, Railway station, …, Airport | Airport, Bus Station, Train Station, Subway, …, Ferry, Light Rail | – |
| Drink & Eat | Catering services | Restaurant, Café, Diner, …, Bakery, Breakfast Spot | Eating and drinking |
| Leisure & Sport | Recreation, Travel | Park, Bar, Gym, Fitness Center, …, Hotel, Movie Theater, General | Leisure and sports |
| Education | Research related, Culture related, School | College, University, …, School, Library | Educational activities |
| *Home* | Housing estate | Home (private), Residential Building | Personal care, …, Household activities |
| *Work* | Work location | Factory, Professional Places | Working and work-related activities |
| Other | Other facilities, Building | Building, City, Bridge, …, Moving Target, Event Space | Other activities, …, Telephone calls |

## 3. Datasets description and processing procedure

This work integrates massive mobile phone data, POI data, and LBSN data to understand the interaction between the inhabitants and the facilities in a big city, Shanghai, China. Call detail records (CDRs) and various data records (XDRs) are two typical mobile phone data used to study the mobility behavior of the population. CDRs collect the time and location of the interactions between users and the base stations when they are making phone calls, sending or receiving text messages. XDRs additionally collect the internet data accessing activities, recording more consecutive mobility traces of users. The XDRs data used for this study are collected from 1,000,000 anonymized users in Shanghai for 2 weeks in January 2014, containing approximately 566 million records in total. On average, each person has 40 records per day. In this work, we focus on the human behavior on weekdays, thus the records of the weekends are eliminated. Each record in XDRs contains an anonymous user ID, the time at the instance of the phone activity, the longitude and latitude of the interacting base station. Fig. 2A-C illustrate the basic statistics of the phone usage behavior in XDRs. Fig. 2A shows the temporal distribution of raw XDRs. Fig. 2B presents the distribution of the number of daily records, suggesting most users have 30 −50 records every day. Fig. 2C presents the distribution of time intervals between two consecutive records of the same user. The XDRs collect most users' locations within a few hours, although some time intervals are large. To filter out the days with too sparse records, we split the entire day into 48 timeslots and eliminate the days if less than 12 timeslots are with records (Widhalm et al., 2015).

The POI dataset adopted in this work contains more than 30,000 POIs in Shanghai (Long, 2016; Jin et al., 2017). The POIs are grouped into 7 categories, including *shopping*, *education*, *leisure & sport*, *drink & eat*, *daily life*, *transport*, and *other*. For each mobile phone user, we keep their *work*, *home* as two additional POI labels. It is noteworthy that, because we do not have access to the full-version check-in data of individual users in Shanghai, we utilize the Foursquare check-in data in Tokyo to derive the visitation pattern of various types of POIs, under the assumption that inhabitants in Shanghai and Tokyo share similar lifestyles. Our proposed method has potential to be further refined with sufficient data available in the same region. The Foursquare data in Tokyo collected over 570,000 check-ins during 10 months from April 2012 to February 2013 (Yang et al., 2014). Each record includes a timestamp, longitude and latitude coordinates, and an activity type.

In addition, owing to the difference in mobility behavior between visitors and local inhabitants, the possible visitors are eliminated in the Foursquare data via removing the users whose timespans between the first and last check-in records are less than two weeks. For the sake of simplicity, we manually match the activity types in Shanghai POIs and Foursquare data with our self-defined types, as illustrated in Table 1.

## 4. Mobile phone data processing and stay labeling

### 4.1. Stay detection with spatial clustering

*Stay* points denote locations where the users have stayed for a while such as schools and offices. Li et al. proposed the stay point detection algorithm that if the distance between a point and its successors is larger than the threshold, they measured the time span between the first point and the last successor (Li et al., 2008). If the time span is larger than the time threshold, the points are labeled *stay*.

In the vast majority of cases, the users interact with the nearest base station. However, when the nearest base station is fully occupied, new connections will be assigned to his/her second nearest station. This reassignment results in the mismatching between users and base stations in space. Therefore, when detecting people's *stay* points, we take such particularity of base station data into account. We first cluster the users' footprints using the DBSCAN algorithm within a certain period, respecting the chronological order, to filter out the sudden changes in location. The footprints belonging to each cluster are replaced by the location of the cluster center. The spatial threshold of DBSCAN is set as 50 *m* in our experiments in Shanghai. After removing the noises in users' trajectories, we identify the significant locations via another DBSCAN algorithm to cluster the active locations ignoring their arrival time and label the clusters as *stay* or *pass-by*. Users engage in activities at *stay* locations and transfer between activities at *pass-by* locations. The stay detection algorithm is shown in Algorithm 1. *Stays* are distinguished from *pass-bys* based on the temporal and spatial thresholds set as 10 *min* and 300 *m* in our experiments in Shanghai. *Stay* locations are identified as the regions where a user has stayed over a certain time interval within one place. By doing so, the points within a certain distance are recognized as the same active significant place. The users' mobility traces are further updated via filling up the significant places, namely the activity

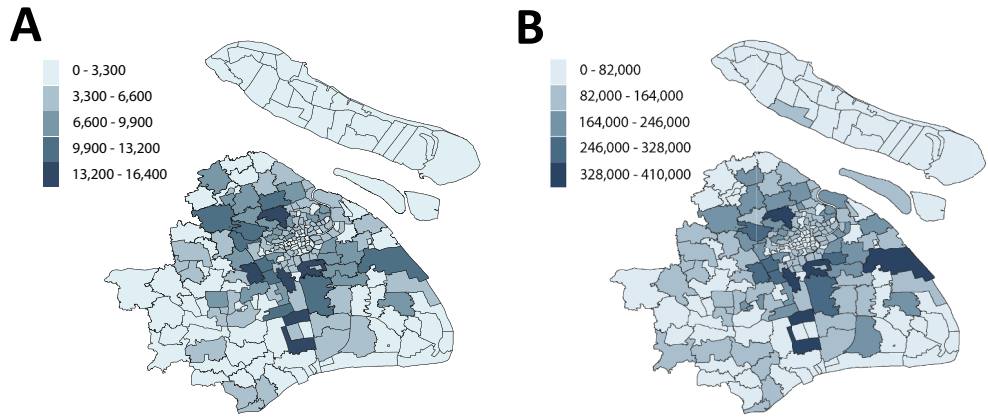

**Fig. 3.** Spatial distribution of home locations of the mobile phone users in XDRs (A) and the population from 6th national census data in 2010 (B) by Jiedao. The users' address in XDRs depends on the inferred home. The shade color from dark to light corresponds to the population from more to less.

chain.

**Algorithm 1:** Stay detection

```
1: L_u∈[], j←1, i←1, C←[]
2: add r_1 to C
3: for i←2 to N do
4:    if (t_i − t_j)⩽10min then    ▷Time interval less than 10 min
5:       add r_i to C
6:    else
7:       add C to L_u
8:       C←[]
9:       j←i
10:      add r_i to C
11: for C in L_u do
12:    C_m←DBSCAN(eps = 0.05, min-sample = 2, C)
13:    for each cluster D in C_m do
           ▷Loops through each cluster in the records within 10 min
14:       the location of cluster D_x, D_y = Med(x_d, y_d)
              ▷Update locations by the medoid of locations in each cluster
15:       update location: x_d, y_d = D_x, D_y
16: L_m = DBSCAN(eps = 0.05, min-sample = 1, L_u)
17: for each cluster U in L_m do
18:    the location of cluster U_x, U_y = Med(x_d, y_d)    ▷Update locations by the
          medoid of locations in each cluster
19:    update location: x_d, y_d = U_x, U_y
20: j←1, i←1
21: for i←1 to N do    ▷Loops through each record of one user
22:    if x_i, y_i = x_j, y_j then
23:       j←j + 1
24:    else
25:       if t_j − t_i⩾10min then    ▷Time interval more than 10 min
26:          la←stay
27:          i←j
28:       else
29:          la←pass-by
30:          i←j
```

We next present the distribution of the number of visited locations per user per day in Fig. 2D. The distribution matches well with a log-normal function with the mean value equal to 4.76, suggesting the users visited nearly 5 places per day on average. As we have derived the stay places for each user, the travel distance between two locations is then calculated with the Euclidean distance. By assuming all residents start their trips from home and back home at the end of the day, we are able to estimate their daily travel distances. Fig. 2E depicts the distribution of users' daily travel distance. It also follows a log-normal distribution with the average travel distance equal to 5.75 km. Fig. 2G shows the spatial distribution of travel flow above 0.01% of total demand during the morning peak hours on one typical weekday.

### 4.2. Stay labelling to identify significant places

People engage in daily activities with certain regularity. Having completed the detection for *stay* and *pass-by* locations, we further label the stays with semantic information, such as *home*, *work*, and *other*. Most research employs simple rules for home detection, which are mainly divided into two categories. Usually, *home* is identified as the location with the most records during the night or the location that has the most records above a predefined threshold (Jain et al., 1999; Tizzoni et al., 2014). In this work, we consider the daily life habits. That is, the inhabitants prefer to stay at home during the night of weekdays and daytime during weekends or stay at offices during working hours on weekdays. Therefore, we assume the frequently visited location from 22:00 to 6:00 every day as her or his *home*. If the user visits the identified *home* less than 30% of the total number of records, we remove this user as she/he probably be a temperate visitor to the city. Thus, the users refer to residents if not specially illustrated in the following context. We compared the spatial distribution of the home locations of the active mobile phone users in XDRs and the population distribution from the 6th national census data in 2010, as shown in Fig. 3. The number of users in XDRs is evenly distributed to the population distribution of Shanghai by Jiedao, reflecting the reliability of the data sources we adopt. Similarly, we label the most frequently visited place between 8:00 and 18:00 on weekdays as *work*. Note that if the identified *work* place is less than 0.5*km* from *home* or visited less than twice a week, the user probably not be a commuter and the stay location is labeled as *other*. The remaining unidentified stay locations are classified as *other*.

Next, according to the availability of the *work* place, we group residents into commuters and non-commuters because of their obviously different mobility behavior. After the types of *stay* places are identified, we can further label their trips into three categories, (i) home-based-work (HBW), representing the trips between home and work places; (ii) home-based-other (HBO), representing the trips between home and other places; (iii) non-home-based (NHB), representing the trips between two other places or between work and other places. Fig. 2F depicts the fraction of hourly trips in these three categories. The peak and trough of HBW travel demand are very pronounced in the curve, and the two peaks correspond to the morning and evening commuting peak hours.

Fig. 4 illustrates the joint distribution of the arrival time and stay duration for *home*, *work*, and *other*. For both commuters and non-commuters, there are two main clusters in *home*. The lower cluster represents the activities at home during the daytime, and the upper cluster indicates the activities during the night. The upper tilts to the lower right, suggesting that the later the user arrives home, the less time they will stay. Compared with commuters, the lower in the *home* activities of non-commuters is darker, showing that non-commuters stay much more time at home during the daytime than commuters. In the distribution of *other* activities, we can observe a more evident peak in the night for commuters than non-commuters, confirming the fact that commuters only can spend a long-time on other activities at night. In contrast, the *other* activities of non-commuters increase during the day and reduce at night. As to the *work* activity, from Fig. 4A, we can observe that *work* can be split into three hotspots. The lower two hotspots represent the work with two time intervals, morning and afternoon because some works leave their workplace for a break in the noon. The upper hotspot indicates that the workers stay in the work place all day for about 8 h.

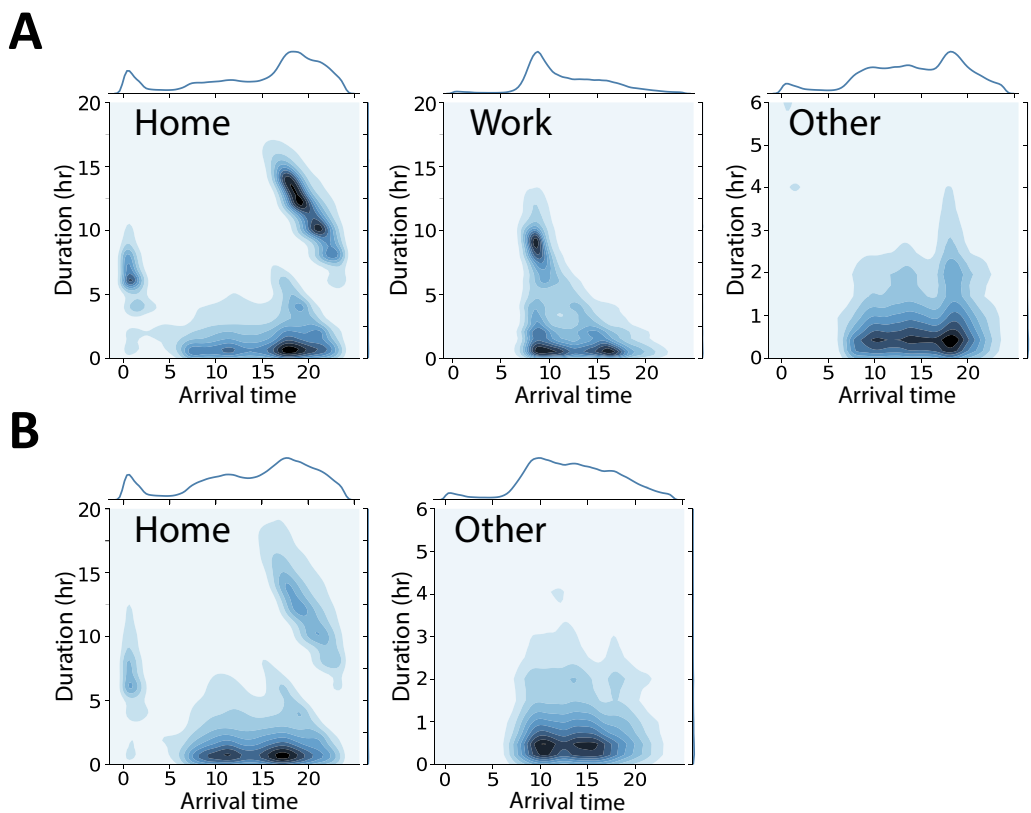

**Fig. 4.** Joint distribution plot of the stay duration and arrival time of commuters (A) and non-commuters (B) for inferred activities from the XDRs data, including *home*, *work* and *other*. Most of the stay duration at *home* of the commuters falls between 10 to 15 h. The typical stay duration at *work* is around 8 h. While for the non-commuters, they often stay at *home* for a short time during the daytime.

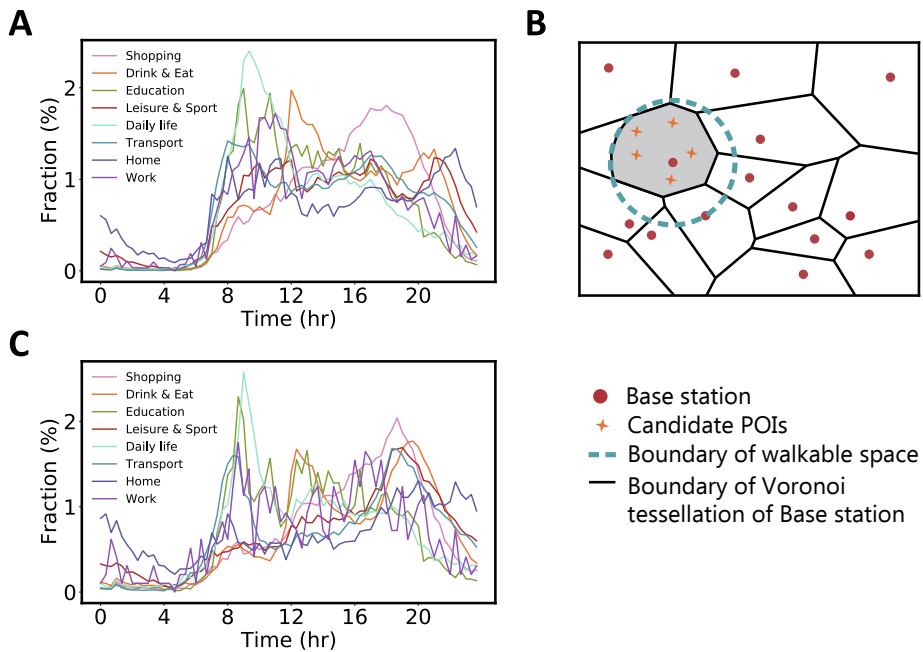

**Fig. 5. Traveler's activity reconstruction with Bayesian inference and the LBSN data.** (A) Distribution of the arrival time of various types of activity LBSN data, which are collected from Foursquare check-ins in Tokyo. The temporal granularity is 15 min. (B) Selection of the candidate POIs for the Bayesian inference framework. The solid black polygons present the Voronoi diagram of the base stations (red dots). The blue dotted circle refers to the walkable space of the users. The grey shaded area shows the selected region, in which all of the POIs (orange stars) can be selected by the users visiting the target base station. (C) Distribution of the arrival time of the different types of activities estimated by coupling the XDRs and LBSN data. It is worth noting that *home* and *work* activities are inferred from daily habits, rather than Bayesian inference model.

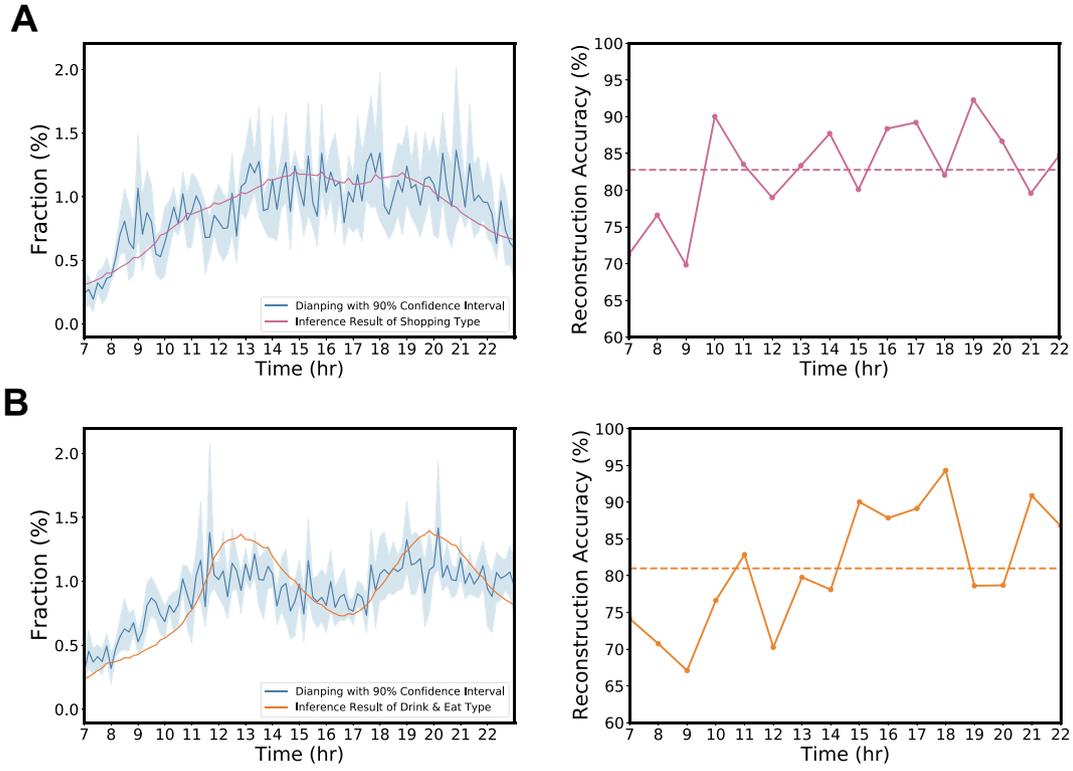

**Fig. 6.** Validation of inferred activities from Bayesian model with real-world data collected from Dianping for *shopping* (**A**) and *drink & eat* (**B**) type between **7:00 and 22:00.** We compared the time distribution of these two types of activities with a time granularity of 10 min and calculated the reconstruction accuracy as one minus mean absolute percentage error (MAPE) per hour. The dashed line represents the mean value of reconstruction accuracy. Our results demonstrate a high level of consistency and reconstruction accuracy (around 80%) with real-world human behavior.

## 5. Activity inference with Bayesian estimation

Due to the rhythm of inhabitants' daily lives, the temporal access patterns differ across activity types. For example, the eating activities have evident noon and evening peak hours compared to other activities. On the other side, because of the coarse localization of mobile phone data, we can only limit the possibly visited POIs within a polygon based on the base station locations. Therefore, the choice of POI needs to consider both the POI density within the polygon and the access patterns of different activity types. Here, we propose a Bayesian activity inference model that relies on the distribution of POIs and the temporal pattern of different activity types extracted from the LBSN data.

### 5.1. Bayesian activity inference model

We convert the activity type inference problem as estimating the probability of various activity types at the given arrival time and place. The possible visited POIs, a.k.a. candidate POIs, are selected based on the Voronoi polygon of the base station. In practice, people display different visitation patterns to various types of POIs. For instance, restaurants always have evident peaks during the evening hours, while cultural facilities always attract more visitors on weekends. Fig. 5A presents the temporal change of the attractiveness of different types of POIs in the LBSN data. As expected, the vast majority of the activities happen during the daytime; *daily life*, *education*, and *work* activities display a high attractiveness during the morning hours; *drink & eat* and *home* show two evident peaks during the morning and evening hours; *leisure & sport* and *shopping* show evident peak during the evening hours. On the other side, *daily life*, *education*, and *work* activities have a much larger probability to be visited than the *leisure & sport* and *shopping* activities during the morning peak hours.

With consideration of such temporal inhomogeneity of visitation patterns, our target is to estimate the conditional probability of various

POIs locating around the users' destinations. As shown in Fig. 5B, we first select the candidate POIs (the orange stars) in the Voronoi polygon of the base station visited by the user. To avoid selecting some distant POIs locating in the polygons near the boundary of Shanghai, we also eliminate the ones outside the 900 meters buffer of the base station (the blue dashed circle). Let $D = (c, t)$ denotes the activity with the proportion $c$ of the candidate POIs' types and the time $t$. Next, given a set of POIs, we set the visit probability of the POI type $O_i$ as $p(O_i|c, t)$, where $i = 1, 2, \ldots, 7$, representing each activity type except *home* and *work*. According to the Bayes theorem, the probability of visiting a certain POIs type is

$$p(O_i|c, t) = \frac{p(t|O_i, c)p(O_i|c)p(c)}{p(c, t)}. \tag{1}$$

It is assumed that the time and location of activity are independent of each other, then $p(t|O_i, c) = p(t|O_i)$. The probability of visiting a certain type of POIs can be simplified to:

$$p(O_i|c, t) = \frac{p(t|O_i)p(O_i|c)}{p(t)}. \tag{2}$$

where $p(O_i|c)$ represents the probability of selecting a certain POI type in the case of the given distribution of all POI types in the activity region, $p(t|O_i)$ represents the temporal probability to visit a certain POI type, and $p(t)$ is the probability of accessing the timeslot $t$. Since we use a time granularity of 10 min, there are a total of 144 timeslots in a day, i.e. $p(t) = \frac{1}{144}$. The inferred activity type $O_i^*$ can be further written as:

$$p(O_i|c, t) \propto p(t|O_i)p(O_i|c) \tag{3}$$

$$O_i^* = \text{argmax}_i p(t|O_i)p(O_i|c) \tag{4}$$

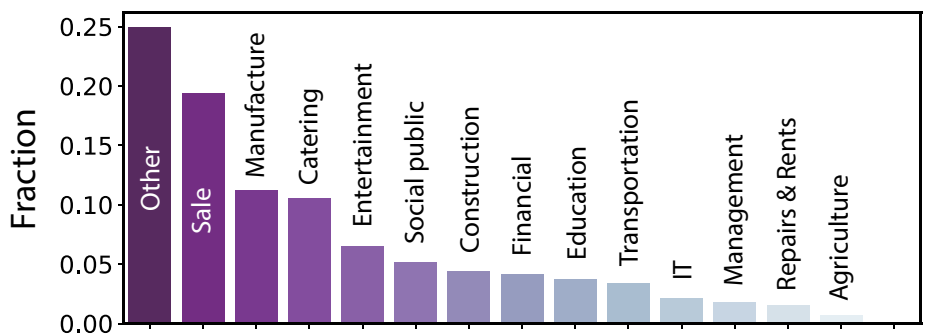

**Fig. 7. Distribution of professions of all commuters after Bayesian inference.** Sale is the top profession in Shanghai. Manufacture is the second top, closely followed by catering.

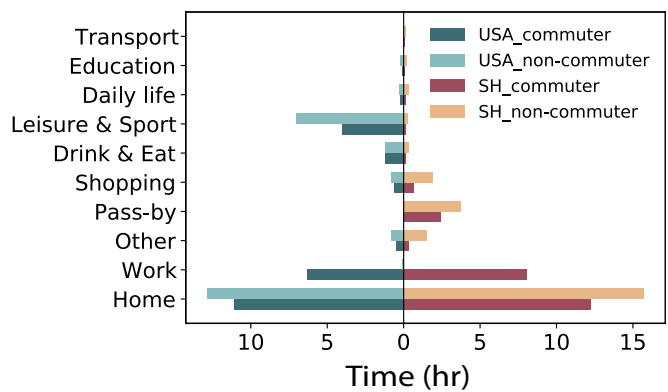

**Fig. 8. Comparison of stay duration of each activity for the commuters and non-commuters in the XDRs and the American Time Use Survey (ATUS).** Note that there is no *Pass-by* in the ATUS. The left and right panels show the time use in U.S. and Shanghai, respectively. The non-commuters in Shanghai stay at home for a longer time than people in the U.S. The commuters in Shanghai stay longer time at both home and work than people in the U.S. In contrast, people in the U.S. spend more time for *Leisure&Sport* than our XDR users in Shanghai.

### 5.2. Validation of Bayesian activity inference results

Fig. 5C illustrates the temporal distribution of each activity type after inferring the user activity of XDRs via combining LBSN data and POIs. As can be seen, the temporal tendency of each activity after Bayesian inference keeps in alignment with the original temporal distribution of Foursquare check-ins (Fig. 5A). For example, the *drink & eat* activities have similar three peaks at 8:30, 12:30, and 20:30, corresponding to breakfast, lunch, and dinner time, respectively.

We validate our inferred human activities by comparing them against actual data collected from Dianping. Similar to Yelp or Foursquare, Dianping allows users to "check-in" at businesses they visit, reflecting the real activity type of the user at that time. The check-in data we use includes anonymized user ID, the time and date of the check-in and the type of the business. Our primary objective is not only to accurately predict individual activity types but, more importantly, to reconstruct overall patterns at a citywide level and better understand human activity. On the other side, there is no activity records in our mobile phone datasets. Therefore, here we focus on validating whether the patterns we obtain are consistent with real-world behavior with Dianping data.

Specifically, we collected users' check-in records for *shopping* and *drink & eat* respectively every January from 2014 to 2020, as our mobile phone data were collected in January. We then compare the time distribution of these two types of activities with our reconstructed results to see if our model can infer the true patterns of human activity. When calculating the time distribution of our reconstructed activities, we consider the entire duration of the mobile user's stay instead of just arrival time since users may check in at any point during their stays.

In Fig. 6, we find that our reconstructed results are consistent with real-world human behaviors. For *shopping* activities, the fraction gradually increases in the morning hours and maintains a high level in the afternoon and evening, gradually decreasing after 20:00. For *drink & eat* activities, there are two obvious peaks in the check-in data at around 12:00 and 19:00, and our reconstruction results successfully captures this pattern. Further, we calculate the mean absolute percentage error (MAPE) per hour respectively and use one minus MAPE as a measure of reconstruction accuracy. Both types of activities demonstrate high reconstruction accuracy, around 80% in Fig. 6. Overall, our model can effectively capture the patterns of human activity on a large scale.

## 6. Understanding the inferred activity patterns

After inferring and validating the activity types in the user trajectories from mobile phone data, we conduct further analysis on the reconstruction results to gain more insights into human activity behavior. One such insight is the identification of spatio-temporal patterns inherent in the inferred activities. Specifically, we utilize the spatial information of *work* activities to infer the occupations of commuters, and use the temporal information to deduce the time use of Shanghai inhabitants. Additionally, we employ an LDA topic model to explore varied activity sequences and gain an understanding of different lifestyles of Shanghai residents.

### 6.1. Spatio-temporal patterns of activities

#### 6.1.1. Profession distribution of commuters

The POIs and previously inferred *work* locations enable us to infer the commuters' professions. Understanding the distribution of professions facilitates further understanding of the economic structure and urban land use. With reference to the International Standard Industrial Classification of All Economic Activities (ISIC) and the industry classification in the Third Shanghai Economic Census, we classify POIs into 14 groups corresponding to 14 categories of occupations. For example, IT includes computer programming, the Internet and communications, etc; the social public includes health-related, government, environment-related, etc. Due to the limited information provided by POIs and XDRs, the possible occupations of commuters are only based on geographic location and semantic. The candidate POIs for the profession are chosen with the Voronoi of base stations belonging to the *work* location. The conditional probability of profession inference can be inferred by the categories' proportion of the candidate POIs. When the POIs around the *work* locations are hazy or missed, the profession may be difficult to infer. Occupation is inferred as other in two cases, the selected poi of the *work* location is ambiguous or its category is other. As Fig. 7 presents the proportions of commuters in different professions, the largest proportion of commuters is actually unknown occupations. We compared the inferred results with the third Economic Census of Shanghai in 2013 (Bureau of Statistics of Shanghai, 2015). Among commuters whose occupations are inferred clearly, those in sales occupations accounted for the largest proportion (close to 20%), consistent with the economic census, where the sales also accounted for the largest proportion (exceeding 15%). People working in social public accounted for 5.16% and those working in construction accounted for 4.4%, which is consistent with those in the economic census. In the Shanghai Economic Census, 4.4% of people work in social public, and 3.9% of people work in real estate.

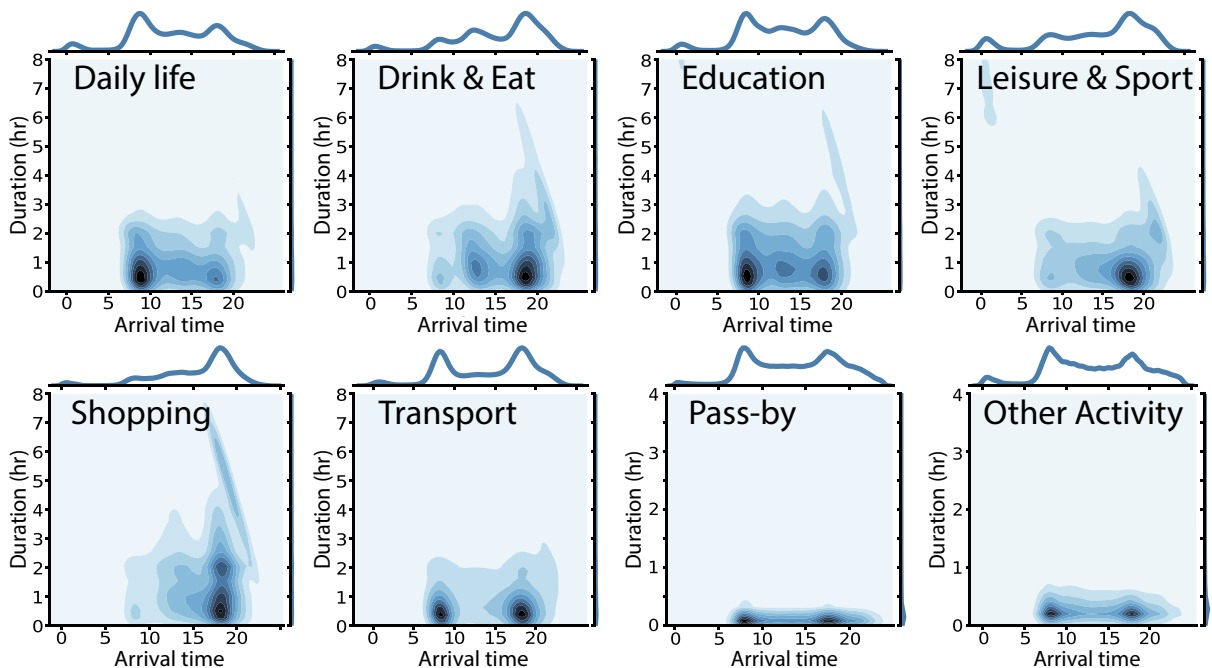

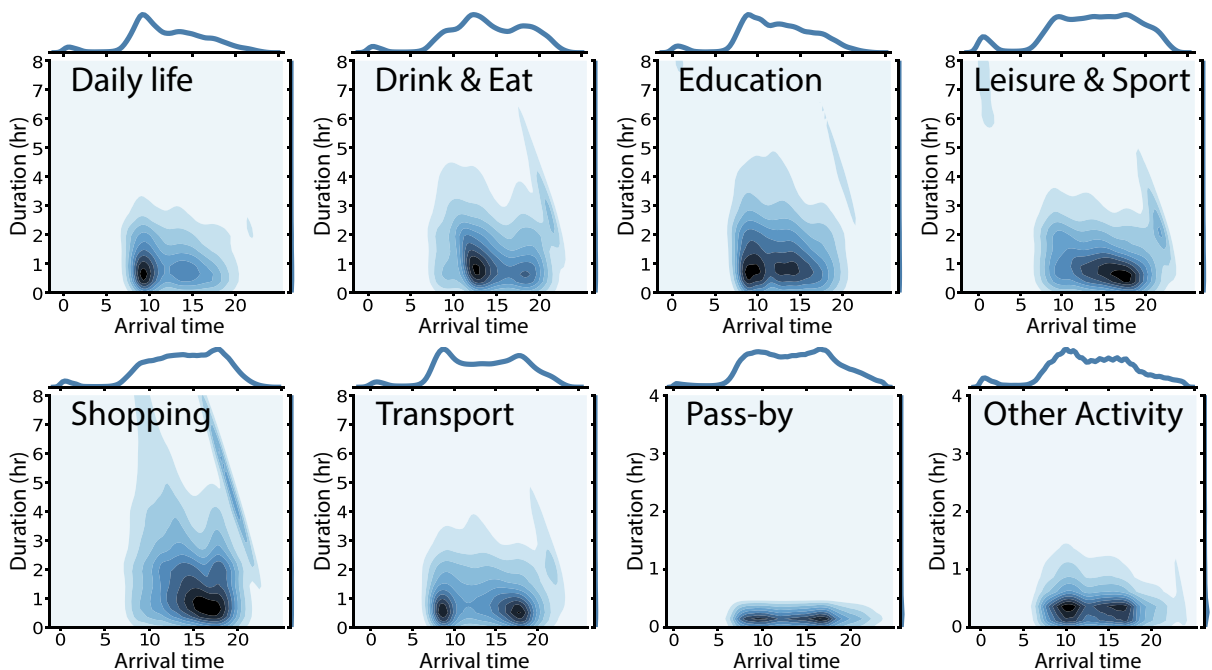

**Fig. 9. Joint distribution plot of duration and arrival time of the commuters (A) and non-commuters (B) for various activity types.** The distributions of arrival and duration agree well with the reality. For example, *drink & eat*, *leisure & sport* and *shopping* often start around 7:00 PM for commuters, while these activities often start during the daytime for non-commuters.

*6.1.2. Daily time use analysis*

Based on the inferred activity types, we also compare the average time use distribution with the American Time Use Survey (ATUS, (U.S. Bureau of Labor Statistics, 2009)), as shown in Fig. 8. Covering 105 million U.S. households aged 15 and older, ATUS provides nationally representative estimates of how, where, and with whom Americans spend their time. The left and right panels of Fig. 8 present the time use of different activities from ATUS and Shanghai, respectively. The darker and light colors present the time use behavior of commuters and non-commuters, respectively. The comparison shows that people in Shanghai spend more time at home and work than U.S. people.

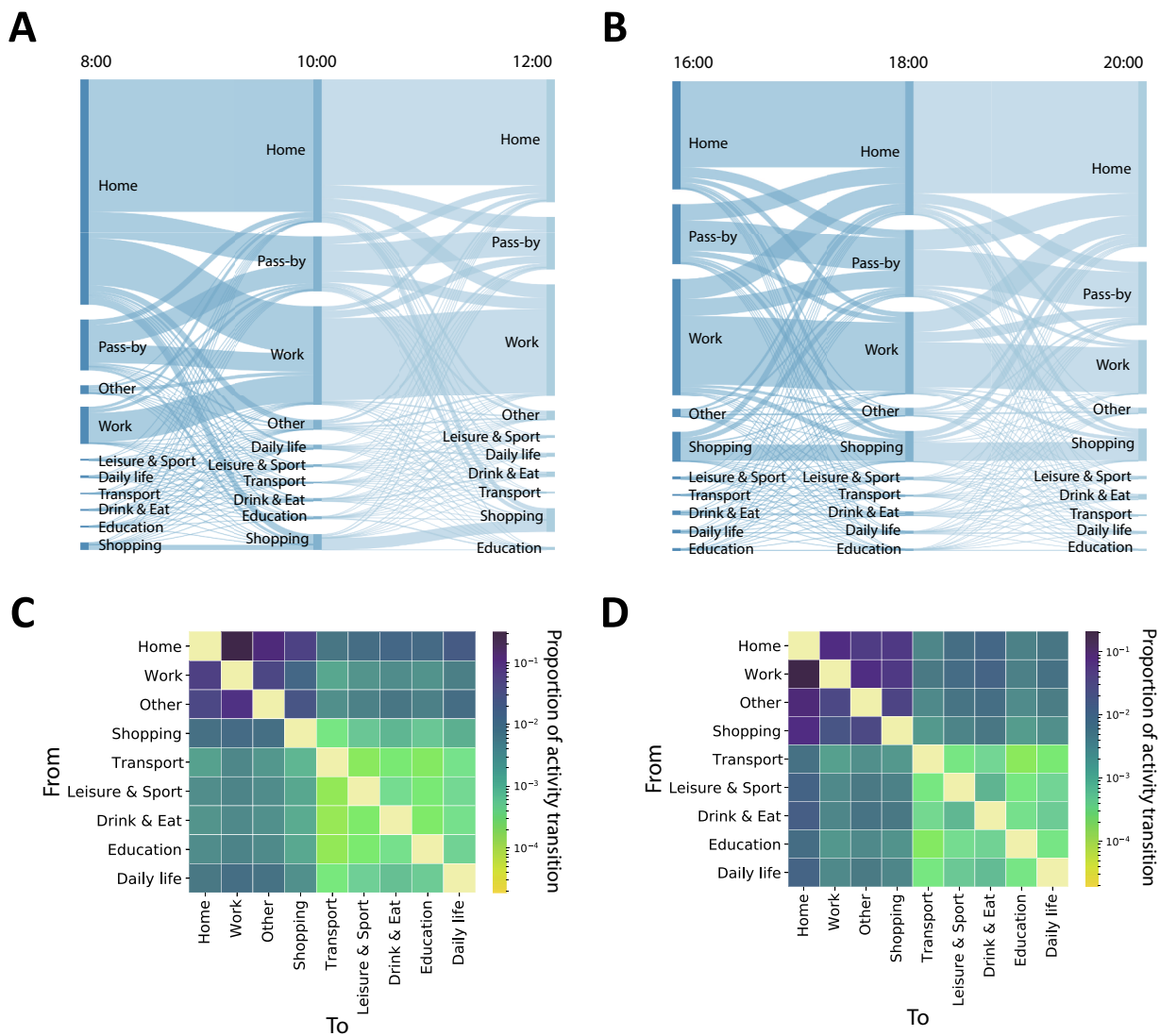

**Fig. 10. Transfers of population flow between different activities.** (A-B) Population flow between different activities during the morning peak period (8:00 AM-10:00 AM-12:00 AM), and evening peak period (4:00 PM-6:00 PM-8:00 PM). (C-D) Transfer matrices between different activities from 8:00 AM to 12:00 PM, and from 4:00 PM to 8:00 PM.

Commuters in Shanghai work nearly 8 h per day on average, while people in the U.S. work around 6 h per day. We also observe that people in Shanghai spend less time on entertainment and dining activities than Americans while spending the opposite amount of time on family activities. This indicates that Shanghai people prefer to engage in family activities than go out for socializing or entertainment during their free time.

To compare the temporal characteristics of various activities, in Fig. 9, we show the joint distribution of the duration and arrival time for different activities for commuters and non-commuters, respectively. Most activities of commuters concentrate in some peak hours during one day. For example, commuters' participation in *leisure & sport* or *shopping* activities peaks at around 19:00, while non-commuters have frequently activities from 12:00 to 20:00. Compared with commuters, non-commuters start earlier and last longer when engaging in entertainment activities. Commuters' *drink & eat* activities have three clusters in time, with peak hours at 8:00, 12:00, and 18:00, respectively corresponding to three meal times. However, non-commuters' only has two clusters with peak periods at 12:00 and 18:00, suggesting that commuters prefer to have breakfast outside than non-commuters. The *transport* activity of commuters has two clusters with marked peaks, representing the trips to work in the morning and from work in the afternoon. In contrast, the arrival time of non-commuters' ranges from morning to night. Non-commuters start *education* activities mainly around 9:00 and 13:00, matching real-life habits, while commuters start *education* activities mainly at 9:00 and 18:00. The reason for commuters having an evening start time could be due to the fact that after work, some may attend educational institutions to learn interest courses such as dancing, musical instruments, etc. The availability of such courses in the evening follows their work schedule.

We next examine the transition between different activities of the inhabitants, as presented in Fig. 10. Fig. 10A and C show the activity flow and transition matrix during the morning peak period respectively. Fig. 10A presents the average activity flow from 8:00 to 12:00 on weekdays. The proportion of residents visiting the workplace from 8:00 to 10:00 increases significantly. At noon, a small fraction of workers return home. The labels on the left side in Fig. 10C indicates the activity users transit from and those on the bottom indicate the activity users transit to. The activity with the highest percentage of departures is home and the most significant transition is from home to work during the morning peak hours. Fig. 10B and D show the activity flow and transition matrix during the evening peak period, respectively. At 16:00, most people are at work. A part of people leave work and move to other places at 18:00. At 20:00, the probability of going home increases significantly, but a few people are still at work. The most popular transition in the corresponding transfer matrix is from work to home.

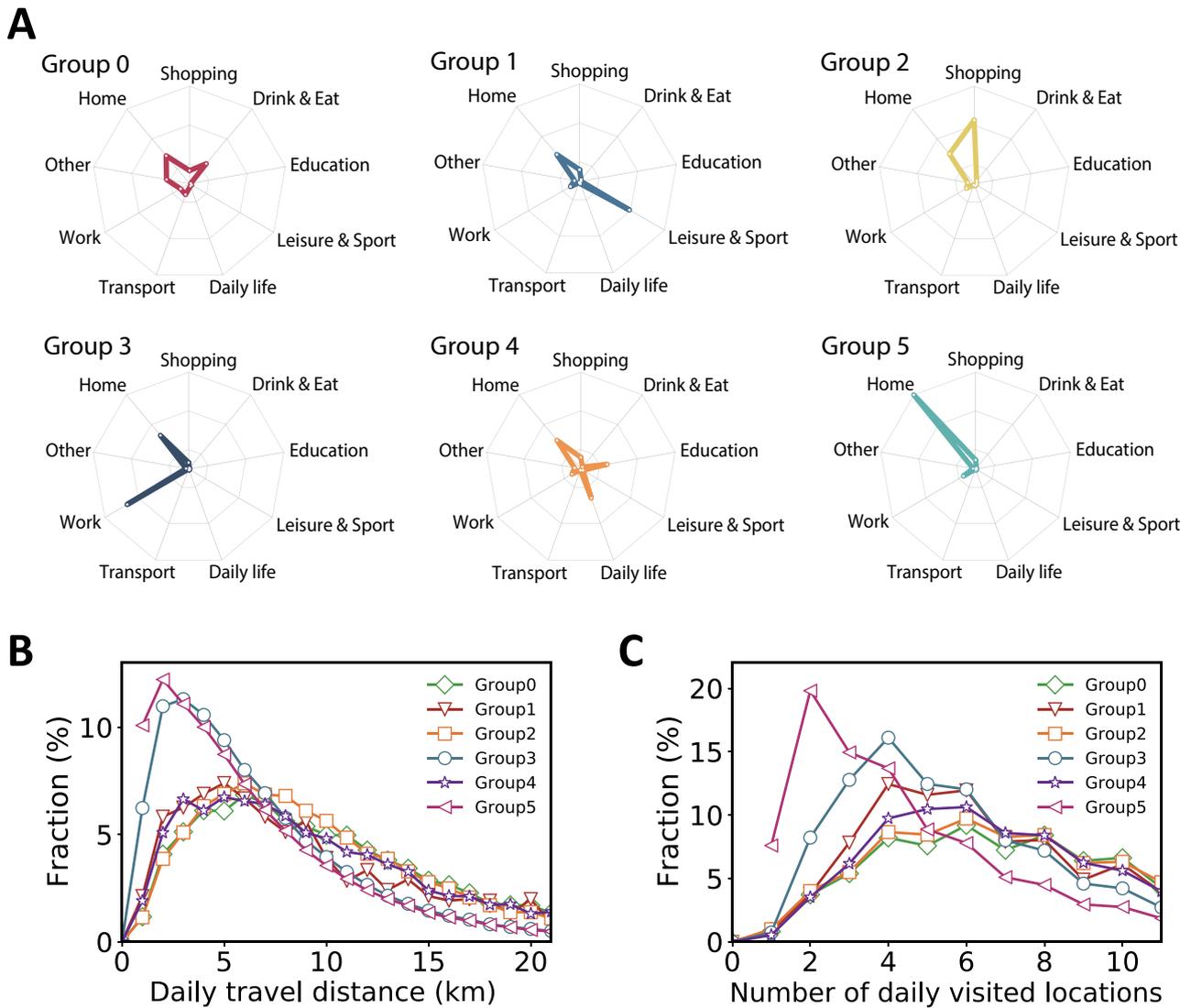

**Fig. 11. The activity and trajectory characteristics of different groups of users after the classification by LDA.** (A) Comparison of the proportion of different types of activities in different groups. (B) Comparison of daily travel distances between users in different groups. (C) Comparison of the number of locations visited by users in different groups.

### 6.2. Different patterns of activity chain with LDA model

#### 6.2.1. LDA model and gibbs sampling

The sequence of individual activity, a.k.a, activity chain, is an important feature of daily lives, reflecting daily behaviors and habits. Further insight into the residents' behavior characteristics provides a new opportunity for enriching the urban activity model. To this end, we use a topic model to classify users into different groups based on their activity chains. Topic models are unsupervised machine learning models used in natural language processing (NLP). These models are usually used to discover topics that cannot be observed directly in texts. LDA (Blei et al., 2003) is one of the most widely used, assuming that each document includes a fixed number of implicit topics, and the topics are represented by the word distribution. LDA aggregates the same words in a large number of documents, gives the topics in each document in the form of the probability, and finally classifies the documents according to the topic distribution.

We regard the activity sequence $S$ of each user as a document and the activity type $A$ in each time interval as a word. Thus, the category $Z$ is equivalent to a topic. Suppose there are $M$ users and $V$ activity types in total, user $m$ has $N_m$ activity types in the activity sequence $s_m$, where the topic of the $n$-th activity $a_{mn}$ is $z_{mn}$. In the LDA model, the activity type in each category obeys a multinomial distribution sampled from the Dirichlet prior distribution of the parameter $\vec{\beta}$, and the category in each time use sequence follows the same distribution with the parameter $\vec{\alpha}$. For each user's activity sequence $s_m$, its category distribution is:

$$\theta_m = Dirichlet(\vec{\alpha}),$$

then,

$$z_{mn} = multi(\theta_m),$$

Similarly, for each category $k$, its activity type distribution is:

$$\varphi_k = Dirichlet(\vec{\beta}),$$

then,

$$a_{mn} = multi(\varphi_{z_{mn}}).$$

The probability that an activity type is initialized to $d$ can be calculated

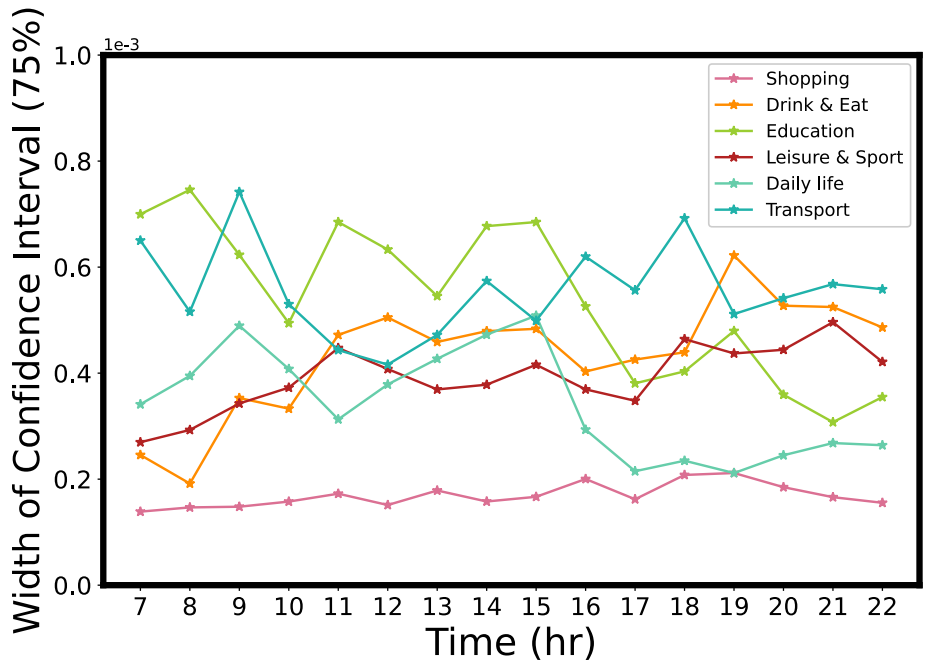

**Fig. 12.** Width of 75% confidence interval for different activity types from 7:00 to 22:00. *Shopping* activities have the narrowest confidence intervals, indicating stable and consistent inferred results. During the night, *daily life* and *education* activities exhibit narrower confidence intervals compared to daytime. However, this trend is reversed for *drink & eat* activities. The confidence interval width for *leisure & sport* activities remains constant throughout the day.

as:

$$p(a_{mn} = d | \overrightarrow{\theta_m}, \Phi) = \sum_{k=1}^{K} p(a_{mn} = d | \overrightarrow{\varphi_k}) p(z_{mn} = k | \theta_m),$$

The likelihood function of the entire activity sequence is performed as the following:

$$\mathscr{L} = \prod_{m=1}^{M} p(\overrightarrow{a_m} | \overrightarrow{\theta_m}, \varphi) = \prod_{m=1}^{M} \prod_{n=1}^{N_m} p(a_{mn} | \overrightarrow{\theta_m}, \varphi)$$

Next, Gibbs Sampling (Wang, 2008) is used to solve the hidden variables $z_{mn}$, $\overrightarrow{\theta_m}$, $\overrightarrow{\varphi_k}$ of LDA. Initially, each activity in the sequence is assigned a topic $z^0$ randomly. We count the number of activity types $d$ appearing under topic $z$ and the number of categories $z$ in the activity sequence of user $m$. Then the fully conditional posterior distribution $p(z_i | Z_{-i}, A)$ can be obtained, where $Z_{-i}$ denotes all $z_j$'s with $j \neq i$ and $i, j$ are the activity indices in the activity sequence. With the probability distribution of the current activity type in all categories, a new category $z^1$ is assigned to the activity based on the probability. The same procedure is repeated until the category distribution $\overrightarrow{\theta_m}$ of each activity sequence and the activity type distribution $\overrightarrow{\varphi_k}$ of each category converge. And we can get the required parameters and the category of each activity. The probability formula of the category of current activity is:

$$p(z_i = k | \overrightarrow{Z_{-i}}, \overrightarrow{A}) \propto \frac{n_{k,\neg i}^t + \beta_t}{\sum_{t=1}^{V} n_{k,\neg i}^t} (n_{m,\neg i}^k + \alpha_k),$$

where $n_k^t$ is count of activity $t$ in all activities assigned topic $k$ and $n_m^k$ is count of activity in sequence $m$ assigned $k$. The subscript $\neg i$ means that $a_i$ and $z_i$ are not included in the calculation. And the parameter calculation formula is performed as the following:

$$\varphi_{kt} = \frac{n_k^t + \beta_t}{\sum_{t=1}^{V} n_k^t + \beta_t},$$

$$\theta_{mk} = \frac{n_m^k + \alpha_k}{\sum_{k=1}^{K} n_m^k + \alpha_k}.$$

#### 6.2.2. Activity trajectory patterns analysis

To model the users' activity chains, we first flatten the user's activity sequence with the equal time interval of half an hour based on the inferred activity types. The time range of the activity sequence is limited between 6:00 to 22:00 to select key activities, regardless of the large number of household activities at midnight. After comprehensive consideration of coherence within the topic, differences between topics in Appendix B and human judgment in Fig. 11A, we eventually assign the active mobile phone users into six groups and analyze the difference of their behavior in two aspects: daily travel distance and the number of daily visited locations.

Fig. 11A illustrates the activity distributions of various groups. Among the six groups, people in group 0 participate in various activities in a balanced and active way and are labeled as active users. Group 1 is marked as leisure-led with the largest proportion of recreational activities in addition to *home* activities. Similarly, group 2 labeled as shopping-led has the largest proportion of *shopping* activities and people in group 3 are labeled as work-led users with the largest proportion of *work* activities. People in group 4 prefer to participate in education-related and daily life activities. They may include students and educators and are labeled as education-led users. Group 5 is purely home-led. As shown in Fig. 11B and C, the daily activity distances of work-led users and education-led users are relatively small in six groups, whereas the daily activity distances of entertainment-led users and active users are relatively large. Fig. 11C compares the average daily visit locations of different groups. The work-led and home-led users visit fewer locations per day, while active residents visit more.

### 7. Discussion and conclusion

With the advent of the big data era, people share their real-time locations to gain various types of services, offering an unprecedented opportunity to understand the population's mobility behavior and their interaction with the urban functions. With the most large coverage in population, mobile phone data have been widely used to model human mobility. However, coarsely localizing users with base stations results in the difficulty of inferring the activity details. Many research analyzed the LBSN data that have semantic information of their own to infer activity types, and few studies explored human activity with mobile phone data. Most of them focus only on the related semantic information (POIs) such as the attractiveness and the available time. We propose to couple the mobile phone data with the POIs and LBSN data together. In addition to the relationship between the geographical environment and POIs, we also consider the attractiveness change of activities in different time-slots. POIs data provides the geolocation of facilities and the LBSN data provides the temporal visitation pattern to various types of POIs. To this end, we devise a Bayesian model to assign each POI a possibility to visit for each trip in the mobile phone data. Taking Shanghai as an example, we first extract the trip chains of 1,000,000 anonymized users during two weeks with the XDR data. Combined with the 570,000 check-in items and more than 30,000 POIs, we are able to infer the activity type of the active mobile phone users in Shanghai. The validation results with a real-world check-in data show that our model has a high level of consistency and reconstruction accuracy (around 80%) with real-world human behavior.

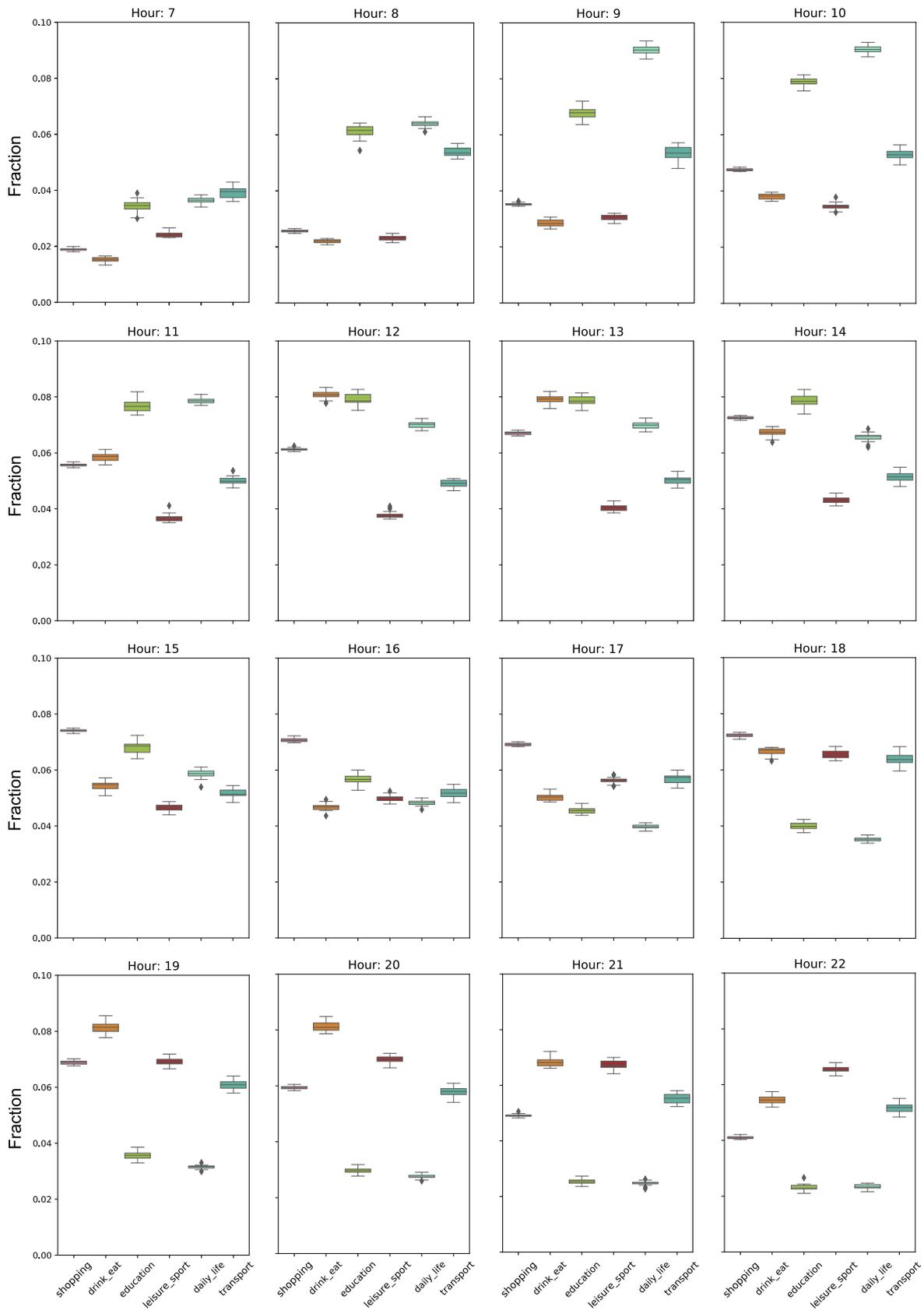

**Fig. 14. Box plot of time distribution from 7:00 to 22:00 for each activity type.** We randomly select 20 subsets from the mobile data, with each subset comprising 20% of the total users, and conduct the Bayesian inference model on each subset.

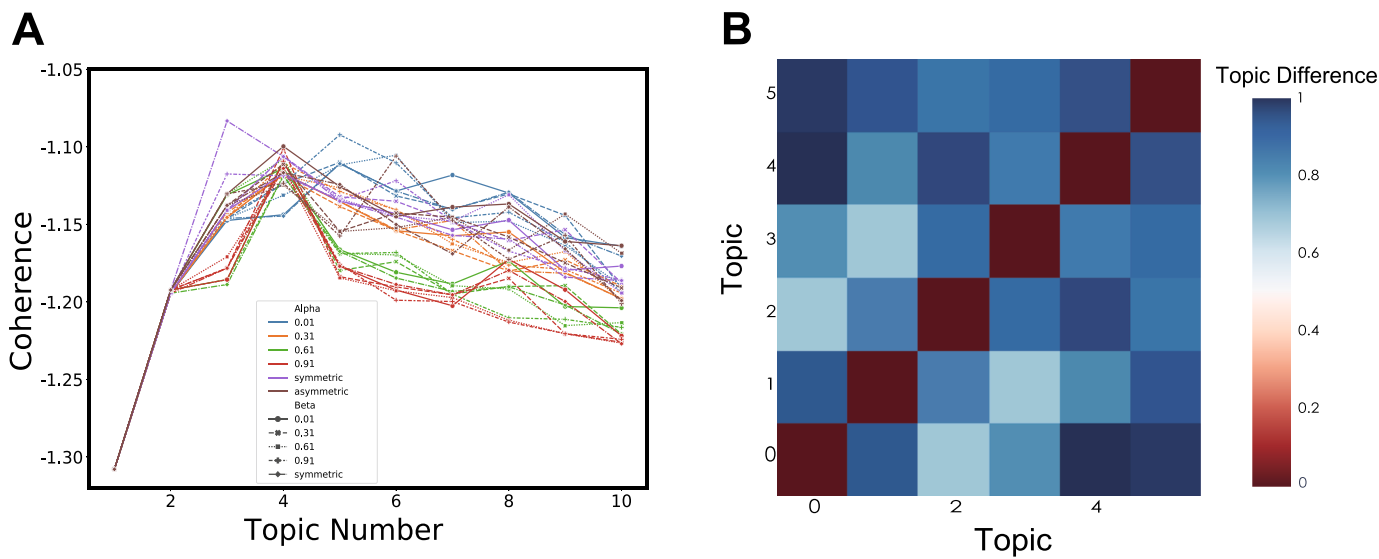

**Fig. 13.** Analysis of LDA model. (A) Sensitivity analysis on the LDA model by altering three hyperparameters, including $\alpha$, $\beta$ and the number of topics. When $\alpha$ or $\beta$ is symmetric, it means using a fixed symmetric prior of $1.0/$ num_topics. When $\alpha$ is asymmetric, it uses a fixed normalized asymmetric prior of $1.0/$ (topic_index $+$ sqrt(num_topics)). When the number of topics is within the range of 4 to 7, the coherence score is relatively high. (B) Measurement of differences between pairwise topics. The bluer the grid on non-diagonal lines, the better, because this indicates a greater degree of differentiation and fewer intersections between the two topics. Our result demonstrates a high level of dissimilarity between each topic pair.

After inferring the activity types in the users' trajectories on a large scale, we proceed to analyse the reconstructed outcomes from different aspects. Firstly, we analyze the temporal characteristics of each activity type and the average daily time use. We observe evident differences between commuters and non-commuters. For example, the *drink & eat* of commuters has three clear peak periods, whereas non-commuters have only two concentrated clusters in time; the *leisure & sport* and *shopping* of commuters start in the evening whereas the start time of non-commuters is distributed throughout the day. Next, via comparing the time use of the active mobile phone users in Shanghai with the American Time Use Survey, we find that the inhabitants in Shanghai work about one hour and a half longer on average than those in the United States. To understand the activities of different groups of people, we adopted a topic model, LDA, to group the users into six categories, reflecting their various lifestyles. We find that the work-led and home-led users have shorter travel distances and fewer visited locations on average per day and the leisure-led and active users show the opposite.

It is noteworthy that, due to the lack of check-in data of population in Shanghai, we leverage the publicly available Foursquare data in Tokyo (Yang et al., 2014) as a proxy for the visitation pattern to different categories of POIs. We believe this assumption can be feasible as the proposed Bayesian method only utilizes the visitation frequency per 10 min in our framework (as shown in Fig. 5A), and the lifestyle of population in Shanghai and Tokyo is quite similar. We will quantitatively compare the similarity if data in Shanghai is available in future. In addition, although we have considered the temporal pattern of the visitation to different types of facilities, future research can be extended by improving the accuracy of user activity type inference, such as considering the interaction between different activity types and the capacity of different facilities or focusing on the study of specific activity schedules, such as commuting which are important trips during workdays.

## Appendix A. Performance of Bayesian inference model

To demonstrate the performance of our Bayesian inference model, we randomly select 20 subsets from the original data, each containing 20% of mobile users, conduct the Bayesian inference model on each subset, and calculate confidence intervals for each activity type. A wider confidence interval indicates that there is more uncertainty in the model's reconstruction results. Since *home* and *work* activity types are derived from daily life habits in 4.2 rather than inferred from Bayesian model, we mainly focus on the remaining activity types.

In Fig. 12 and Fig. 14, we can see that regardless of whether the fraction of *shopping* activities is high or low, it has the narrowest confidence intervals, indicating that the inferred results for *shopping* activities are the most stable. Other than that, *daily life* and *education* activities have narrower confidence intervals at night than during the day while *drink & eat* activities are the opposite in Fig. 12. The confidence interval width for *transport* activities is similar throughout the day.

## Appendix B. Sensitivity analysis of LDA model

We perform a sensitivity analysis on the LDA model by altering its hyperparameters, including $\alpha$, $\beta$, and the number of topics, in order to evaluate their impacts on the model's performance. The number of topics is a crucial hyperparameter in LDA, as tailoring this parameter can determine the balance between fit and generalization. Hyperparameters $\alpha$ and $\beta$ regulate the sparsity of document and topic distributions. To conduct our study, we examined six different $\alpha$ values (0.001, 0.031, 0.061, 0.091, 'symmetric', 'asymmetric'), five $\beta$ values (0.001, 0.031, 0.061, 0.091, 'symmetric'), and ten distinct sets of topic numbers (ranging from 1 to 10). The LDA model's performance is evaluated by the UMASS coherence score (Röder et al., 2015) which measures the degree of semantic similarity between pairs of words within a chosen topic. A higher coherence score indicates that the topics generated by the LDA model have a higher degree of interpretability and represent meaningful themes within the corpus.

Our findings, as presented in Fig. 13A, indicate that despite varying $\alpha$ and $\beta$ values, the trend of coherence score changes is basically consistent. Specifically, when the number of topics is relatively low, coherence scores are similarly low. Once the number of topics exceeds 4, the change in coherence score tends to stabilize. But when the number of topics is too large, the coherence value actually decreases. Based on these results, as well as our own subjective assessment in Fig. 11A, we determine that a classification result of six topics produces a favorable LDA outcome.

Furthermore, in addition to measuring the coherence score within the topic, we also use Hellinger distance (Arora et al., 2013) to measure the differences between the six types of topics. The result is demonstrated in Fig. 13B, with the colors of the grid representing the degree of correlation between the vertical and horizontal topics. Our results show a high degree of difference between each topic pair since all grid colors except diagonal are blue.